\documentclass[12pt, preprint]{aastex}
\begin{document}
\title{Ultraheavy Element Enrichment in Impulsive Solar Flares}
\author{David Eichler }
\affil{$^1$ Physics Department, Ben-Gurion University, Be'er-Sheva 84105, Israel}
\email{eichler.david@gmail.com}
\begin{abstract}
{Particle acceleration by cascading Alfven wave turbulence was suggested (Eichler, 1979b) as being responsible for energetic particle populations in $^3$He-rich solar flares. In particular, it was noted that the damping of the turbulence by the tail of the particle distribution in rigidity naturally leads to dramatic enhancement of pre-accelerated species - as $^3$He is posited to be -  and superheavy elements.  The subsequent detection of large enrichment of ultraheavies, relative to iron, has apparently confirmed this prediction, lending  support to the original idea. It is shown here that this picture could be somewhat sharpened  by progress in understanding the 3-dimensional geometrical details of cascading Alfven turbulence (Sridhar and Goldreich, 1995).  The mechanism may be relevant in other astrophysical environments where the source of turbulence is non-magnetic,  such as clusters of galaxies.}
\end{abstract}

\section{Introduction}
  Non-thermal particle tails in particle energy spectra are evidence for a selective acceleration process that chooses to accelerate some particles but not most. The question to be asked of any acceleration model is which particles are selected and which are not. Shock acceleration, while it clearly selects only a minority of particles to be accelerated to very higher energies  [presumably the ones that already have a higher than average energy (Eichler, 1979c)], is remarkably equitable in regard to the different ion species. While there is modest enhancement of heavy elements, much of the differences in abundances from normal solar composition [e.g. correlation with first ionization potential (FIP)], can be attributed to "chemical" adjustment of the underlying composition of the thermal plasma.  (For example, any  electromagnetic separation of ions from neutrals could differentiate low and high FIP species from each other in partially ionized plasma. In the case of anomalous cosmic rays, the bias towards high FIP species is traced to the "pre-injection" they enjoy when the solar UV suddenly ionizes them after they have penetrated the solar wind in their neutral state. When this happens, they become the most energetic particle species in the solar wind plasma, and are more easily picked up by the shock acceleration mechanism.)

 In impulsive solar energetic particle (SEP) events, $^3$He  is often dramatically enriched relative to $^4$He nuclei. The $^3$He nuclei are believed to also receive an initial boost  - due to their resonance with ion-cyclotron waves (Fisk 1978) - but $^3$He-rich flares are not associated with shocks, and are generally smaller than shock-associated flares. Rather, it was proposed (Eichler 1979b, hereafter E79) that the $^3$He ions continue their acceleration by cyclotron damping turbulence as it cascades to smaller scales. Because the $^3$He nuclei have received a sufficient amount of pre-acceleration, they  have  larger gyroradii than protons,  (besides having a slightly smaller charge to mass ratio, which would be true of any heavy ion species). They need only a factor of $\sim 3$ increase in velocity to comfortably meet this criterion. At the scales where thermal or epithermal ions damp Alfven wave turbulence as it cascades to smaller spatial scales, the turbulence is a steeply decreasing function of wavenumber k, so that the $^3$He ions, which cyclotron resonate at the lower k, resonate with turbulence of  a much larger amplitude  and are preferentially accelerated. It was noted in E79 that extremely heavy elements would also experience highly preferential acceleration because they are not completely stripped. For example Fe$^{+10Q_{10}}$ has a gyroradius, at a given velocity, of $5.6Q_{10}$ times that of ions, and $2.8Q_{10}$  times that of $^4$He. An ultraheavy with $A\sim 200$ would have at most slightly more charge than iron, and would have a gyroradius that is more than 10 times that of protons of the same velocity. (Note that long wavelength magnetosonic waves tend to bring all ion species to the same velocity in a collisionless plasma [Eichler, 1979a], so the assumption of equal thermal velocities is reasonable.) The basic idea in E79 has since been confirmed in simulations (Miller, 1998). Observations (Mason et al., 2004) indicate that ultraheavies are indeed significantly enriched, even relative to iron, in $^3$He-rich flares. On the other hand, there are also energetic protons; apparently protons are not completely shut out of the acceleration process.

Formal expressions for the rates of particle diffusion in pitch angle and in momentum have been given by Kennel and Engelman (1966) and also by Lee and Volk,  (1975), Bogdan et al (1991) and Schlickeiser and  Achatz  (1993). The discussion below will be self-contained, however, with emphasis on particle diffusion in energy space. It will be assumed that particle velocities greatly exceed the Alfven velocity, so that the energy gain rate is slower than the pitch angle diffusion rate, and pitch angle distribution is therefore nearly isotropic.

The theory of cascading Alfven wave turbulence was revolutionized by Sridhar and Goldreich (1994), who noted that in the limit of sharp resonance the three wave process (symbolized by  ${\bf k_1 +k_2 \rightarrow k_3},\,  \omega_1 + \omega_2 \rightarrow \omega_3$)  invoked by Kraichnan (1966), the basis for the previously popular picture of cascading turbulence, does not  in fact take place. They further noted that, when interacting waves 1 and 2 are oppositely propagating, the four wave interaction $\bf{k_1 +k_2 \rightarrow k_3 +k_4},\,  \omega_1 + \omega_2 \rightarrow \omega_3 +\omega_4$ leaves  the parallel components of ${\bf k_1}$ and ${\bf k_2}$ unchanged, i.e. $k_{1,\parallel} = k_{3,\parallel},\, k_{2,\parallel} = k_{4,\parallel}$, and they conjectured on this basis that the turbulence would cascade more quickly in the perpendicular direction of $\bf k$-space, leading to tubular Alfven turbulence with $k_{\perp} \gg k_{\parallel}$.    

In this paper, in sections 2 and 3  [see also Chandran (2000)],  it is noted that the tubular feature of  Alfven turbulence   would significantly affect the nature of the resonant cyclotron particle acceleration  because a particle whose parallel motion is in resonance with the wave motion would, for most pitch angles, pass through many perpendicular wavelengths, leading to a rapidly changing direction of the electric field that accelerates the particle. This causes the acceleration to be less efficient for low rigidity particles.  On the other hand, it is also  noted here that the acceleration becomes even more selective in favor of high rigidity particles and therefore favors  partially ionized heavy elements, and especially ultraheavy elements. It is shown that the acceleration time is constant in energy, leading to exponential growth in the energy content of superthermal particles, and with a faster growth time for partially ionized heavy ions.   In sections 4 and 5, comparison with observations is made, and it is argued that observations of impulsive solar flares is consistent both with dramatic enhancement of high rigidity ion species and with low overall efficiency, so that cascading turbulence remains an attractive  candidate for impulsive solar energetic particle events.

 \section{Basic Idea}

 The mathematical illustration of selective acceleration given in E79 was  simplified.  It was assumed that turbulent energy cascades monotonically to high k at a rate $dk/dt$ that is a given function U(k) of k, which would give a cascade time of $k/{dk\over dt} =  k/U$. In fact, the cascading is a non-linear process and its rate is more correctly expressed in proportion to the dimensionless amplitude of the turbulence ${\cal W}(k)k/(B^2/8\pi)$. However, Sridhar and Goldreich (1995) have argued that the level of turbulence adjusts itself so that the cascade time is $t_{cascade}\sim 1/v_A k_{\parallel}$, in which case  U(k) can indeed be expressed independently of  ${\cal W} (k) dk$.

 The condition for cyclotron resonance is that $k_{\parallel}v_{\parallel} -\omega - n\omega_c=0$. For non-relativistic particles, this is a condition on $v_{\parallel} \sim [E/m]^{1/2}$. Let us focus on the n=1  resonance, and for simplicity, on any single species of ions. Highly super-Alfvenic  particles, $v_{\parallel}\gg v_A= k_{\parallel}^{-1} \omega$  of velocity $v \sim k_{\parallel}^{-1} \omega_c$ satisfy the resonance condition so the spectral energy density can be expressed as a function of the parallel velocity of the particles that resonate with it, i.e.  ${\cal W} (k) dk \equiv W( v_{ \parallel})dv_{ \parallel}$. The cascade time at a given wavenumber is then naturally expressed in terms of the velocity of the resonant particles as $t_{cascade} = v_{\parallel}/v_{A}\omega_c$.

 The  average diffusion of particles in energy space due to mode k is given by \\${ D_{EE} = <[\Delta E]^2/\Delta t>\eta_{\parallel}\eta_{\perp}\sim \, <(dE/dt)^2>\eta_{\parallel}\eta_{\perp}\Delta t}$ where $\Delta E$ is the correlated energy change over a correlation time $\Delta t$, and where $ \eta_{\parallel}\eta_{\perp}$ is the fraction of particles that remain in phase with the wave over the correlation time. The energy change is due to electric fields in the observer frame acting along the particle velocity. For Alfven waves, the electric field  ${\bf \cal E}$  is perpendicular to the magnetic field, so  $dE/dt = ev_{\perp}\cdot {\cal E}$ and the phase correlation time,  $\tau_c$,  of the wave over which resonant particles are affected is, according to  Sridhar and Goldreich, $\tau_c \sim 1/k_{\parallel }v_{A}$. The electric field  ${\cal E}_k$ due to the {\bf k}th mode is given by ${\cal E}_k = {\bf u_k  \times  B} \simeq u_k B_o$. The fraction $\eta_{ \parallel}\sim(\Delta v/v)$ of particles at $v_{\parallel } \sim \omega_c/k_{ \parallel}$ that resonate with the wave is determined by the criterion that they gain or  lose a phase of less than $\pi/2$ over the correlation time $\tau_c$ (i.e. $\Delta v \tau_c = \pi/2k_{\parallel}$), and, assuming $\tau_{c} \sim \tau_{cascade}$, $\eta_{ \parallel}\sim v_A/v_{\parallel}$.  Of those that are within the parallel resonance,   the fraction  $\eta_{\perp} = 1- \cos\theta $ of particles whose pitch angle $\theta$ is small enough that their gyration does not carry them across a distance of more than $1/2\pi$ of a perpendicular wavelength $2\pi/k_{\perp}$, i.e. those obeying $k_{\perp} r_g \sim {k_{\perp} v_{\perp}\over \omega_c } \lesssim 1$, can be small when $k_{\perp} \gg k_{\parallel}$, which lends added importance to the modes $k_{\perp} \lesssim k_{\parallel}$.
Finally we can write

 \begin{equation}
 D_{EE}  
  = e^2 B_o^2  \sum_k u_k^2 v_{\perp}^2{\eta_{\perp}}/c^2k_{\parallel }v_{ \parallel}\simeq  \sum_k m  u_k^2 mv_{\perp}^2{\eta_{\perp}}\omega_c.
 \end{equation}
 Below we will be concerned with nearly isotropic particle distributions, but with anisotropic wave spectra, so we have assumed in the above that parallel and perpendicular velocities are typically of the same order for most of the particles, and, in equation (1), $v_{\perp}^2$ can be replaced in any pitch angle average by $2v^2/3 =  2v_{\parallel}^2$. The factor $\eta_{\perp}$ essentially selects out the modes $k_{\perp} \lesssim k_{\parallel}$ in the sum, as will be seen below.
 
The acceleration time $\tau_a = E^2/D_{EE}$  is typically  longer than the cascade time  $\tau_{cas}=v/v_A\omega_c$ by the factor $(v/v_{\perp})^2 (v/u)^2(v_A/v  \eta_{\perp})$. The power  per thermal proton mass absorbed by a resonant particle of velocity $v_{\parallel}$ and perpendicular velocity $v_{\perp}$, $v_{\perp}^2 u^2 n(v_{\parallel},v_{\perp})\omega_c/v^2 n_{th}$, exceeds the  cascade power per thermal proton  mass, $u^2v_A\omega_c/v_{\parallel}$  only if the ratio   $n(v_{\parallel}, v_{\perp})/n_{th}$ of the number density of resonant particles, $n((v_{\parallel},v_{\perp})$, to the total, $n_{th}$, exceeds
 $(v_A/v)(v/v_{\perp})^2 /\eta_{\perp}$. This condition cannot be met unless the fraction of resonant particles exceeds $v_A/v$.  Unless the plasma is weakly magnetized or the resonant particles are already highly superthermal, it does not appear that the cascading turbulence could be damped until it cascades down nearly to scales where it could damp on the tail of the distribution of typical thermal ions. 

 Let us  then assume  the turbulence cascade proceeds as in Sridhar and Goldreich (1995) without  too much energy absorption   by resonant  particles.  The 3-dimensional energy spectrum, by their arguments, is given by ${\cal W}(k_{\perp},k_{\parallel})\propto k_{\perp}^{-10/3}g(k_{\perp}^{2/3}k_m^{1/3}/k_{\parallel})\equiv k_{\parallel}^{-5}f(k_{\perp}^{2/3}k_m^{1/3}/k_{\parallel})$ where $k_m$ is the maximum wavenumber at which the turbulence is reasonably isotropic, and g and f are functions  that decline sharply beyond an argument of order unity.  This means that the wave energy $\int \int{\cal W}(k_{\perp},k_{\parallel})2\pi k_{\perp} dk_{\perp} dk_{\parallel}\sim W k_{\parallel}^3$ that is resonant with particles of energy $E \sim mv^2/2$ increases as ${\cal W}k_{\parallel}^3 \propto k_{\parallel}^{-2} \propto v_{\parallel}^2 \propto E$. In deriving this result we have taken  the integral over $k_{\perp}$  from $k_{\perp}=0$ only out to to $k_{\perp} =  k_{\parallel}$. The reason for this is that waves with $k_{\perp}\gg k_{\parallel}$ do not effectively accelerate the particles because their characteristic gyroradius $v_{\parallel}/\omega_c$, as defined by the resonance condition, cuts across many perpendicular wavelengths for pitch angles of order unity. (Alternatively, we could have considered particles with very small pitch angle, $\theta \le k_{\parallel}/k_{\perp}\ll 1$, but they  comprise only a small fraction of the total number of particles at a given energy [i.e. $\eta_{\perp} \ll 1$], and also absorb  less energy because they have such a small $v_{\perp}$. So in the end, most of the energy is absorbed by particles with larger pitch angles [$\eta_{\perp} \sim 1$] from the more parallel propagating waves.)

 The energy diffusion rate $D_{EE}$ then, by equation (1), scales as $EW(k_{\parallel},0)k_{\parallel}^3 \omega_c \propto E k_{\parallel}^{-2} \omega_c $, where $ k_{\parallel} = \omega_c/v_{\parallel } \sim  \omega_c/v$.  The acceleration time $E^2/D_{EE}$ is then independent of energy within any given ion species.  This means that the typical particle energy grows exponentially with time in the absence of escape.

 Significantly, it also implies that the acceleration time for partially ionized heavy elements is {\it shorter} than for protons  because, although their cyclotron frequency is smaller by a factor Q/M (where Q and M are in units of protons charge and proton mass respectively), the factor $k_{\parallel}^{-2}$ is larger by $(M/Q)^2$.  Altogether, the acceleration rate of a partially ionized heavy atom is $M/Q$ times that of a proton. Because the energy grows exponentially in time, the enhancements of ultraheavy elements at high energies can be huge.

\section{A Formal Derivation of the Acceleration Rate}
 We now present a more formal calculation justifying the above arguments:  The energy diffusion rate is given by
 \begin{equation}
 D_{EE}({\bf v})=<\Delta E(t)>^2/t =  <\int_0^t q{\bf v\cdot {\cal E}}{(\bf x[t^{\prime}]},t^{\prime}) dt^{\prime}><\int_0^t q {\bf v \cdot {\cal E}(x[t^{\prime \prime}]},t^{\prime \prime}) dt^{\prime \prime}>/t
 \end{equation}
 where
 \begin{equation}
 {\bf x[t]} = {\bf x_o} + v_{\parallel} t {\bf \hat z} + r_g\sin(\omega_c t + \phi_o){\bf \hat x}+ r_g\cos(\omega_c t + \phi_o){\bf \hat y},
  \end{equation}
   \begin{equation}
 \bf v(x[t]) = v_{\parallel}  {\bf \hat z} + r_g\omega_c\cos(\omega_c t + \phi_o){\bf \hat x}- r_g \omega_c \sin(\omega_c t + \phi_o){\bf \hat y},
  \end{equation}
  \begin{eqnarray}
  {\cal E}(x,t)={\it Re}\int\int \int {\cal E}_{\bf k} e^{i({\bf k\cdot x -\omega t +\phi_k(t))} }d^3k \\
  ={\it Re}\int\int \int {\bf {-u({\bf k})}\over c} \times {\bf B_o} e^{i({\bf k\cdot x -\omega t +\phi_k(t))} }d^3k
 \end{eqnarray}
and

\begin{eqnarray}
{\bf v \cdot {\cal E}_k} = B_o ( u_x(\bf k) v_y -u_y({\bf k}) v_x)e^{i({\bf k\cdot x -\omega t +\phi_k(t))}}/c \\ = B_o r_g \omega_c [u_x (k) \cos(\omega_c t + \phi_o)+ u_y(k)\sin(\omega_c t + \phi_o)]e^{i({\bf k\cdot x -\omega t +\phi_k(t))}}
\end{eqnarray}

We assume  that the Alfven turbulence is axisymmetric around $\bf \hat z$  so that we can without loss of generality consider an Alfven mode with $\bf k$  in the zx plane. In the Alfven mode, $u\parallel {\bf \hat y}$, and
\begin{eqnarray}
{\bf v \cdot {\cal E}_k} = B_o [u_y({\bf k}) v_x)e^{i({\bf k\cdot x} -\omega t +\phi_k(t)]}/c \\= B_o r_g \omega_c [u_x ({\bf k}) \cos(\omega_c t + \phi_o)]e^{i({\bf k}\cdot{\bf x} -\omega t +\phi_k(t))}/c.
\end{eqnarray}
Using the identities
\begin{equation}
e^{iz\sin\phi} =  \sum_{n=-\infty}^{\infty}J_n(z)e^{in\phi},\end{equation} 
and 
  \begin{equation}{2n\over z}J_n(z)=J_{n-1}(z)+J_{n+1}(z),
\end{equation} 
 we can write
\begin{eqnarray}
{\bf v \cdot {\cal E}_k} = B_o r_g \omega_c u({\bf k})
\cos(\omega_c t +\phi_o)]
e^{i({ k_{\parallel}v_{\parallel}}t+k_{\perp}r_g sin(\omega_ct+\phi_o) -\omega t +\phi_k(t))+i{\bf k\cdot x_o}}/c\\
{\bf v \cdot {\cal E}_k}
={1\over2} B_o r_g \omega_c u({\bf k})
[e^{i(\omega_c t +\phi_o)]} + e^{-i(\omega_c t +\phi_o)}]
e^{i({ k_{\parallel}v_{\parallel}}t+k_{\perp}r_g sin(\omega_ct+\phi_o) -\omega t +\phi_k(t))+i{\bf k\cdot x_o}}/c\\
= {1\over2} B_o r_g \omega_c u(k)\sum_{n=-\infty}^{\infty}[J_{n-1}(k_{\perp}r_g) + J_{n+1}(k_{\perp}r_g)]
e^{i[({\bf k_{\parallel}v_{\parallel}}
+n\omega_c -\omega )t +n\phi_o +\phi_k(t)]+i{\bf k\cdot x_o})
}/c\\
=  B_o  \omega_c u(k)k_{\perp}^{-1}\sum_{-\infty}^{\infty} n J_n(k_{\perp} r_g )
e^{i({\bf k_{\parallel}v_{\parallel}}
+n\omega_c -\omega )t +n\phi_o+\phi_k(t))+i{\bf k\cdot x_o})
}/c
\end{eqnarray}

We invoke axisymmetry of $u({\bf k})$ around the z direction and  write

\begin{equation}
\int \int \int { u^2}({\bf k}) e^{i({\bf k\cdot x }-\omega t +\phi_k(t)) }d^3k = \int \int u^2(k_{\perp},k_z) e^{i({\bf k\cdot x }-\omega t +\phi_k(t)) }2\pi k_{\perp} dk_{\perp}dk_{\parallel}
\end{equation}
whence

\begin{eqnarray}
 D_{EE}({\bf v})t &=& \int dk_{\parallel}\int 2\pi k_{\perp}dk_{\perp} \nonumber <\int_0^t d( t^{\prime}+ t^{\prime \prime})/2 \int_0^t d( t^{\prime}- t^{\prime \prime})q^2  \omega_{c}^2  {u^2({ k}) B_o^2\over k_{\perp}^2 c^2 } e^{i[({\bf k_{\parallel}v_{\parallel}}
 -\omega )(t^{\prime}- t^{\prime \prime} ) + \phi_{\bf k}(t^{\prime})-\phi_{\bf k}(t^{\prime\prime})] }\nonumber \\  
&&  \big(\sum_{-\infty}^{\infty} n^2 J_n^2(k_{\perp} r_g )  + \sum_{-\infty}^{\infty} \sum_{m\neq n} nm J_n( k_{\perp} r_g )J_m (k_{\perp}r_g)   
e^{i n\omega_ct^{\prime\prime}-i m \omega_ct^{\prime} }\big)>
 \end{eqnarray}

 Note that  $ nt^{\prime\prime}-mt^{\prime}=(n-m)t^{\prime\prime} + m(t^{\prime\prime}-t^{\prime})$ which means that for $n\neq m$, the integrand oscillates over $ t^{\prime\prime}$, and does not contribute significantly to $D_{EE}$. So we henceforth ignore the terms proportional to $J_n( k_{\perp}r_g)J_m(k_{\perp}r_g )$ for $n\neq m$.

 Assuming that  the correlation time of $\phi_{\bf k} (t)$ is small compared to t, the limits of the $ (t^{\prime}- t^{\prime \prime})$  integral can be taken to be $[-\infty, +\infty]$. We can view the term

 \noindent
 $\int_{-\infty}^{+\infty}  d(t^{\prime}- t^{\prime \prime}) e^{i( k_{\parallel} v_{\parallel}+n\omega_c -\omega)(t^{\prime}-t^{\prime \prime}) +i[\phi_{\bf k}(t^{\prime})-\phi_{\bf k}(t^{\prime\prime})] } $
  as a Fourier component of $e^{i[\phi_{\bf k}(t^{\prime})-\phi_{\bf k}(t^{\prime\prime})]}$, and the ensemble average,  $<\int_{-\infty}^{+\infty}  d( t^{\prime}- t^{\prime \prime})e^{i( k_{\parallel} v_{\parallel}+n\omega_c -\omega)(t^{\prime}-t^{\prime \prime}) +i[\phi(t^{\prime})-\phi(t^{\prime\prime})] } >$, as a Fourier component of  $ <e^{i\phi[(t^{\prime})-\phi(t^{\prime\prime})]}>$.

 If the randomization of phases proceeds as diffusion in $\phi_k$ space, then the probability distribution of $[\phi_{\bf k}(t^{\prime})-\phi_{\bf k}(t^{\prime\prime})]$
is given by
\begin{equation}
p[\phi_{\bf k}(t^{\prime})-\phi_{\bf k}(t^{\prime\prime})]=(2\pi)^{-1/2} exp  \left([\phi_{\bf k}(t^{\prime})-\phi_{\bf k}(t^{\prime\prime})]^2/D_{\phi(k), \phi(k)}[t^{\prime}-t^{\prime\prime}] \right),
\end{equation}
then
 \begin{equation}
  <e^{i[\phi(t^{\prime})-\phi(t^{\prime\prime})]}> = \int p[\phi_{\bf k}(t^{\prime})-\phi_{\bf k}(t^{\prime\prime})]e^{i[\phi(t^{\prime})-\phi(t^{\prime\prime})]}d[\phi(t^{\prime})-\phi(t^{\prime\prime})]  = exp[-|\Gamma  (t^{\prime}-t^{\prime\prime})|],
 \end{equation}
 where $\Gamma = D_{\phi(k), \phi(k)}/4$, and
 \begin{equation}
< \int_{-\infty}^{+\infty}  d( t^{\prime}- t^{\prime \prime})
 e^{i(k_{\parallel}v_{\parallel}+n\omega_c -\omega )( t^{\prime}- t^{\prime\prime})+i[\phi(t^{\prime})-\phi(t^{\prime\prime})]}>
 = {2\Gamma\over[(k_{\parallel}v_{\parallel}
+n\omega_c -\omega )^2 + \Gamma^2 ]}
 \end{equation}

 If $\Gamma \sim k_{\parallel}v_{\parallel}$, as posited by Goldreich and Sidhar, then we can make the approximation
 \begin{equation} {\Gamma\over(k_{\parallel}v_{\parallel}
+n\omega_c -\omega )^2   + \Gamma^2 }\simeq \pi \delta( k_{\parallel}v_{\parallel}
+n\omega_c -\omega ). \end{equation}

Using the above, we now write
\begin{equation}
D_{EE}({\bf v}) = 2\pi \int dk_{\parallel}\int 2\pi k_{\perp}dk_{\perp} q^2  \omega_c^2  {u^2({k_{\parallel}, k_{\perp}}) B_o^2\over k_{\perp}^2 c^2 }\\
 \sum_{-\infty}^{\infty} n^2 [J_n^2(k_{\perp} r_g )  \delta({ k_{\parallel}v_{\parallel}}
+n\omega_c -\omega )
\end{equation}

We now simplify further by assuming that $v\gg v_A$, so that $ \omega = k_{\parallel}v_A$  is negligible compared to $ k_{\parallel}v$. Then

\begin{eqnarray}
D_{EE}({\bf v}) \simeq 2\pi v_{\parallel}^{-1}\int 2\pi k_{\perp}dk_{\perp} q^2  \omega_c^2  {B_o^2\over k_{\perp}^2 c^2 }
 \sum_{-\infty}^{\infty} n^2 u^2(n\omega_c/v_{\parallel}, k_{\perp})J_n^2(k_{\perp} v_{\perp}/\omega_c )
\end{eqnarray}

In the theory of  Sidhar and  Goldreich, $u^2(k_{\parallel},k_{\perp})$, which is proportional to the wave energy,  declines steeply for $k_{\perp}\gg k_* \equiv k_{\parallel}^{3/2}l^{1/2}$ where $l$ is the scale at which $u_{\perp}\sim v_A$, i.e. where $\rho u^2(k_{\parallel},k_{\perp} )k_{\parallel}k_{\perp}^2 \sim B_o^2/8\pi$. For $k_{\perp}\le k_{\parallel}^{3/2}l^{1/2}$, the quantity  $u^2$ should be roughly  constant in $k_{\perp}$, as it is established by mode coupling which should equalize the energy among modes, and it should scale as $k_{\parallel}^{-5}$.\footnote{or equivalently $k_{\perp}^{-10/3}$, given  that  the wavenumbers for a given eddy scale as  $k_{\perp}\propto k_{\parallel}^{3/2}$.} The factor $J_n^2(k_{\perp} v_{\perp}/\omega_c ) $, $n\neq 0$, peaks when its argument is of order unity, and declines asymptotically at larger arguments as $k_{\perp}^{-1}$,  so the integrand declines at  large $k_{\perp}$  at least as fast as $k_{\perp}^{-2}$. For resonant ions obeying $v_{\perp}\lesssim v_{\parallel}\sim \omega_c/k_{\parallel}$, the decline in $J_n$ begins at smaller $k_{\perp}$ than the decline in $u^2$, so we can take $u^2(k_{\parallel}, k_{\perp})$ to be approximately constant in $k_{\perp}$ over the significant range of the integral. 

So
\begin{eqnarray}
u^2(l^{-1}, l^{-1})l^{-3}
\simeq v_A^2
\end{eqnarray}
and, for $k_{\perp}\lesssim k_{\parallel}^{3/2}l^{1/2}$,
\begin{eqnarray}
u^2(k_{\parallel}, k_{\perp})
\simeq v_A^2l^{3} |k_{\parallel}l|^{-5}
\label{usquared}
\end{eqnarray}
whence

\begin{eqnarray}
D_{EE}({\bf v}) \simeq 4\pi v_{\parallel}^{-1} q^2  \omega_c^2  {B_o^2\over   c^2 }
 \sum_{1}^{\infty}n^2 \psi_n u^2(n\omega_c/v_{\parallel}, \omega_c/v_{\perp}) \nonumber \\
\simeq  4\pi v_{\parallel}^{-1} q^2  \omega_c^2  {B_o^2\over   c^2 }
 \sum_{1}^{\infty}n^2\psi_n v_A^2 l^3 (v_{\parallel}/n\omega_c l )^5 \nonumber \\
 =  4\pi mv_A^2 mv_{\parallel}^2 [v_{\parallel}/l\omega_c]^2 \omega_c \sum_1^{\infty}n^{-3}\psi_n
 \simeq  4\pi  (mv_{\parallel}^2)^2 [v_A/l\omega_c]^2 \omega_c
 \label{deeb}
\end{eqnarray}
 where $\psi_n \equiv 2\pi \int_0^{\infty}J_n^2(x) d\ln x$.\footnote{ Although this integral would diverge logarithmically for n=0, the n=0 mode is erased by the $n^2$ coefficient.  The integral  is well defined for $n\ge 1$. }
[In contrast to the above, equation (9) of Chandran (2000) assumes that $u^2(k_{\parallel}, k_{\perp}) = 0$ for  $k_{\perp}\lesssim k_{\parallel}^{3/2}l^{1/2}$. This would seem to neglect the most important modes, which obey  $k_{\perp} \sim k_{\parallel}$, and hence  $k_{\perp}\ll k_{\parallel}^{3/2}l^{1/2}$.]

 Equation (\ref{deeb}), valid for $k_{\parallel} \gtrsim 1/l$, implies,  as argued qualitatively in the previous section,  that the acceleration time $t_{acc}$, over which $D_{EE}t_{acc}= E^2$, is constant, so particles subjected to this acceleration gain energy exponentially  in time. As argued qualitatively in the previous section, the acceleration rate at a given velocity is {\it inversely} proportional to the cyclotron frequency $\omega_c$, so that partially ionized heavy ion species are preferentially accelerated.  For the larger spatial scales,  $k_{\parallel} \lesssim 1/l$, on the other hand,  strong, isotropic Kolmogoroff turbulence is probably be the more appropriate description, and  2nd order Fermi acceleration by cyclotron damping would proceed at a rate  $u(r_g)^2/v^2\tau(r_g)$ where $ u(r_g)$ is the turbulent velocity at the scale of $r_g$, and $1/\tau(r_g)$ is the coherent interaction time with an "eddy" and is equal to $\omega_c$.\footnote{We assume that the strong turbulence randomizes the ion's motion on the timescae of $\omega_c^{-1}$.  If the particle's velocity exceeds that of the turbulence, as we may assume for superthermal particles in subsonic turbulence, then this timescale is shorter than that of the fluid motion.}  For a given $\omega_c$, and for the Kolmogoroff spectrum $u(r_g)\propto r_g^{1/3}$, the acceleration rate varies as $r_g^{2/3}\omega_c/v^2  \propto v^{-4/3} \omega_c^{1/3}$. This does {\it not} favor species with velocity or large M/Q. The cascade  model of Goldreich and Sidhar, on the other hand,  gives  an acceleration rate for  test  particles that is {\it inversely} proportional to  $\omega_c$, thus favoring partially ionized ultraheavies among particles over lower Ze/M ions species at a given velocity, in contrast  to isotropic turbulence, in which the acceleration rate of  a test  particle at  a given velocity is proportional to $\omega_c^{1/3}$.

We have not yet taken into account the backreaction of  the cyclotron damping on the turbulence,  which would surely result in the presence of particles that gain energy exponentially with time. This is  considered in the next section.

\section{Backreaction of Particles on the Alfvenic Turbulence}

Alfvenic turbulence without damping  is difficult to solve formally,  and even more difficult when damping is included.  We  may identify several timescales in the problem  associated with   wave-wave and wave-particle interactions on a given spatial scale:  the cascade times $t_{cascade,\perp}$ and $t_{cascade,\parallel}$ in the $k_{\perp}$ and $k_{\parallel}$  directions, respectively,    and the wave damping timescale due to particle acceleration. Goldreich and Sridhar  argued that, at all spatial scales smaller than $l$ but larger than the dissipative range, the strengh of the turbulence reached a level where $t_{cascade,\perp}\sim t_{cascade,\parallel} \sim  1/k_{\parallel} v_{\parallel}$.  They called this approximate equality between parallel and perpendicular cascade times "critical balance".  Their argument is that resonant three-wave coupling  among Alfven waves is forbidden and the cascade in $k_{\parallel}$  can come about only if there is sufficiently strong mode coupling to broaden mode coupling resonance. Cascade  in the parallel direction,  according to this argument,  must "wait"   for perpendicular cascade.  A similar argument could be made  in the reverse  direction -  that  perpendicular cascade  cannot proceed  more rapidly than the time $1/k_{\parallel}v_A$ required  for the waves to feel  the parallel dimension,   because two dimensional turbulence,  having too many conserved quantities of motion,  does not cascade   to high   wavenumbers.

In E79, where the distinction   between perpendicular and parallel cascade was not made, it was argued  that in steady state, (what would be later called) critical balance should exist between the  turbulent cascade time and the damping time, so that just enough turbulence could trickle down to small spatial scales  to resonantly accelerate low energy ions,  just enough to supply the energetic  population that damps the turbulence.  The simplified mathematical  derivation of this is  based on a simple cascade  model in which there is an ordered flow of energy from low k to high k,   according to the spectral  evolution equation
 \begin{equation}
\frac{{\partial\cal W}(k)}{\partial t}
= -\frac{\partial}{\partial k}
\left({{\cal W}(k)}U(k)\right) -{\cal D}(k).
 \end{equation}
 Here  ${\cal W} (k) dk$ is the energy density in turbulence in waves between  the wavenumbers k and k+dk,  $U(k)\sim k/t(k)$  is the k-velocity with which turbulent energy cascades through scale $k$ ( from $\sim k$ to $\sim 2k$ over time t($k$), say).
 $ {\cal D }(k)$ is the rate at which wave energy is damped by the particles. Let us  assume for the purposes of analysis, as is E79, 
that there is a dominant ion species.  Defining
  $W(E)dE \equiv {\cal W}(k)dk$, where E is the ion energy that cyclotron damps waves at wavenumber k; i.e. $k^{-1}=r_g\equiv mcv/eB=(2E/m)^{1/2}$,  we can write

\begin{equation}
\frac{\partial  W(E)}{\partial t}
=  \frac{\partial}{\partial E}
\left[ W(E)U(E)\right] - \left[ \frac{d W(E)}{dt}\right]_{damping}.
\label{dwdt}
\end{equation}
where $E/U(E) \sim  t_{cascade}$,
and the spectral  evolution of the particles obeys
\begin{equation}
{\partial  f(E) \over \partial t} = \frac{\partial}{\partial E} \left[ \frac{dW(E)}{dt}\right]_{damping}
\label{dfdt}
\end{equation}
Now suppose \newline
$ \left[\frac{dW(E)}{dt}\right]_{damping}=W(E)\rm d(E){\partial  f \over \partial E}$,
 where the energy diffusion coefficient, $W(E)\rm d(E)\equiv D_{EE}$,  is written in a form that is manifestly proportional to the wave energy density $W(E)$.  The steady state solution  to equations (\ref{dwdt}) and (\ref{dfdt}) is then 

\begin{equation}
W(E) = U^{-1} P(E_{max})exp\left(-\int_E^{E_{max}} {\rm d}(E^{\prime}){\partial  f \over \partial E^{\prime}}U^{-1}  dE^{\prime}\right)
\label{wofe}
\end{equation}
where $P(E_{max})$ is the turbulent power per unit volume flowing in from  the largest scale that can accommodate the particles before they escape.
and
\begin{equation}
f(E) = \int_E^{E_{max}} {U(E)\over \rm d(E)E} dE
\end{equation}

This solution is time independent and describes  a constant upward flux  $\cal F$ of particles in energy space. Because the exponent in the right hand side of equation (\ref{wofe}) is $\ln E$, the solution describes a depletion of cascade power, $P(E)= W(E)U(E)$, as the wave energy cascades to smaller spatial scales  (where it is resonant with particles of smaller E), that is a fixed fraction of the total; i.e.
\begin{equation}
dlnP(E)/d lnE = 1
\end{equation}
and
\begin{equation}
P(E)=E {\cal F}.
\label{pofe}
\end{equation}

The time-dependent analogue of this solution is, to a fair approximation, the same as in the above, but with $ E_{max}$ replaced by $E_m(t)$,s  where $E_m(t)$ is defined by the condition that the total energy in the particles $\int_0^{E_m(t)}f(E)EdE$ and in the turbulence $\int_{0}^{E_{max}}W(E)dE$ equals the total amount of accumulated energy $P(E_{max})t$ that has been introduced into the system over time t.

 Equation (\ref{pofe}) implies that the damping rate of wave energy by the particles is equal to the cascade rate, i.e. that the damping and cascading are in "critical balance". This follows from the fact that the flux removed from the turbulence at a given spatial scale by the resonant particles of corresponding energy E is proportional to E, so $P(\alpha E) = \alpha P(E)$.  The resonant particle absorb much of the energy at any given scale but they allow just enough to cascade through to smaller scales so that the acceleration of less energetic particles is, in turn,  just enough to replace them as the acceleration moves them to higher energy bins.

 Do we then conclude that, in a 3D description of wave energy cascade, parallel cascade, perpendicular cascade,  and wave damping all proceed at the same rate? Unfortunately there is one complication:  that only waves  with $k_{\perp} \lesssim k_{\parallel}$  are efficiently damped by the  particles,  whereas most of the wave energy, according to Goldreich and Sridhar,  cascades perpendicularly to $k_{\perp} \gtrsim k_{\parallel}$. The possibility must be considered that wave energy can then make an "end run" at $k_{\perp}\gg k_{\parallel}$ around the resonant particles  and penetrate to high wavenumber without losing much energy to the particles.

 There  are, {\it a priori}, at least two possible resolutions of this:

 1) Enough wave energy is backscattered to lower $k_{\perp}$, because of the energy sink there, so that most of it is absorbed by the particles before cascading to high k.   In this case the one dimensional approximation  may retain validity, and the above analysis would apply with $\int W(E)dE =  \int \rho u^2(k_{\perp},k_{\parallel}) dk_{\perp}dk_{\parallel}$, and $U(E) = E k_{\parallel}v_A $. Using equation (\ref{usquared}), we can write
 \begin{equation}
\int W(E)dE = \int \rho u^2(k_{\perp},k_{\parallel}) d^2k_{\perp} dk_{\parallel} \sim u^2(k_{\perp},k_{\parallel}) k_{\parallel} k_{*\perp}^2 \sim   \rho v_A^2/k_{\parallel} l
 \end{equation}
where $k_{*\perp}\equiv (k_{\parallel}l)^{1/2}/l$ is the maximum perpendicular wavenumber of the energy-rich modes at a given $k_{\parallel}$.

 2) At $k_{\parallel} \gg 1/l$, most of the energy cascades as it would in the absence of damping by particles. In this case, the question is how such turbulence eventually damps  when its energy is to be found in the high $k_{\perp}$,  $k_{\perp} \gg  k_{\parallel}$. Does it damp on thermal particles or on non-thermal ones? Electrons or ions?

 For Alfvenic turbulence, it is easy to show that for $k \lesssim \omega_c/c$, most of the current, as measured in the lab frame, is borne by ions.  This means that by the time the turbulence cascades down to the scale of the ion gyroradius, strong magnetic turbulence is essentially heat, as the effective collision time of a typical current-bearing ion is of the order of the gyroperiod. This may be considered the damping scale of the Alfvenic turbulence, unless a population of ions with larger gyroradii coherently cyclotron damps the turbulence at larger  spatial scales.

 We may compare the  rates of cascade $\tau_{cas}^{-1} = k_{\parallel} v_A$ and of cyclotron damping. The analysis above indicates that the  inverse acceleration time for individual ions participating in cyclotron damping is $\tau_{acc}^{-1} \sim (v_A/l\omega_c)^2 \omega_c$, so the particles absorb energy at a rate of  $f(E)E^2\tau_{acc}^{-1} $, and the  wave energy damps on a  timescale of $\tau_{damp}=\tau_{acc} W(E)/f(E)E \simeq {(v_A/l\omega_c)^{-2} \omega_c}^{-1} [\rho v_A^2/f(E) E^2 ][k_{\parallel}l]^{-1} $.  Here we have used equation (\ref{usquared}) to substitute for $W(E)=u^2(k_{\parallel, k\perp})k_{\perp}^2k_{\parallel}$.  The ratio of cascade timescale $\tau_{cas}$ to  damping  time scale is then given by

 \begin{equation}
 {\tau_{cas}\over \tau_{damp}}=[v_A/l\omega_c][f(E) E^2/\rho v_A^2 ].
 \end{equation}

 In situations where the turbulence is powered by a stirring mechanism on spatial scales much larger than a thermal ion gyroradius, the quantity $ (v_A/l\omega_c) $ is much less than unity,  so if  $[f(E) E^2/\rho v_A^2 ]\lesssim 1$, then the turbulence,  even if well mixed by mode coupling into the regime $k_{\parallel}\gtrsim k_{\perp}$, cascades to scales of thermal gyroradius before  being damped by the tail of the thermal distribution.

  On the other hand, the ions at the thermal tail are accelerated exponentially in time, which means their energy content $ f(E) E^2$ grows significantly over a timescale of $(l/v_A)^2 \omega_c$. If the duration  of  the event responsible for the stirring and the collisional timescale are both much longer than this,  and if the particles do not escape before reaching significantly larger energies, then the particle distribution can  evolve to the point where $ {\tau_{cas}\over \tau_{damp}}\gg 1 $, and the turbulence is absorbed before it reaches scales of thermal gyroradii. Physically,  the requirement for this is that the mixing is such than it introduces more turbulent energy density over time than the {\it local} magnetic energy density, so that $[f(E) E^2/\rho v_A^2 ] \gg  l\omega_c/v_A \gg 1$.
  In a solar flare,  which is powered by magnetic energy, this is a significant constraint,  because one expects that the magnetic field energy would be released  and fully dissipated within several Alfven crossing times, and that enough magnetic energy would remain such that the final Alfven velocity would be within an order of magnitude of the original one.

  On the other hand, it is possible that a  region of strong magnetic field annihilation stirs ups the less magnetized neighboring  regions.  Consider, for example, a magnetic arch on the solar surface with a length scale L of $10^8$ cm,  an equipartition scale $l=10^7$ cm, a field strength of 300 G, and an Alfven velocity of $10^8$ cm s$^{-1}$. Then $\omega_c = 10^{6.5}$  and  the acceleration  time is
  $\tau_{acc} \sim [(v_A/l\omega_c)^2 \omega_c ]^{-1} \sim  10^{4.5}$s,  which is too long for a solar flare.
  For a neighboring region, however,
  take the parameters to be $L=10^8 \rm cm$, $B= 3$G, $v_A = 10^6$ cm s$^{-1}$, but let us still assume that the kinetic energy density $U_K$ is comparable to what it is within the strongly magnetized loop, i.e. $10^4$ times the magnetic energy $U_B$ in the less magnetized, surrounding region at scale L.  Assuming the kinetic energy at scale $l$ is, as in the Kolmogoroff spectrum,  $(l/L)^{2/3}$  of what it is at scale L, then  $l =  (U_B/U_K)^{3/2}L = 10^{-6}L = 10^2$ cm. The dimensionless parameter $ (v_A/l\omega_c)^2 \sim 10^{-1}$, and the acceleration timescale is only about 10 gyroperiods, $\lesssim 10^{-3}$ s. This equipartition length $l$, for a thermal velocity of $10^{7.5} v_{7.5}$ cm/s, $ v_{7.5}\lesssim 1$,  is not much greater than a thermal  ion gyroradius, and in fact,  less than the gyroradius of  a partially ionized ultraheavy. So it would then be reasonable to assume that the turbulence at larger scales is quasi-isotropic with $u \gtrsim v_A$, and that the particle acceleration and wave damping for resonant particles at these scales is described by equations (\ref{dwdt}) and (\ref{dfdt}), and that the  anisotropy  predicted by Sridhar and Goldreich does not set in significantly above a thermal ion gyroradius. Given that the  accelerated particles $E\lesssim 1$ MeV/nucleon, have energies only about $10^3$ above thermal energies $\sim 1 $ KeV/nucleon, there are not enough ultraheavies to significantly damp the turbulence, so that we need not consider their back reaction on the turbulence.  The fact that rare ultraheavies are enhanced even relative to iron in impulsive SEP events, despite the likely conclusion that they do not monopolize the energy budget, suggests that the turbulence is a steeper function of $k_{\parallel}$ than a simple Kolmogoroff spectrum. Yet it is possible that {\it all} of the ultraheavies are accelerated well beyond thermal energies. So perhaps a typical impulsive SEP event is properly parametrized somewhere between each of the two extreme examples  given just above. It is also possible that a large fraction of iron-like nuclei and preaccelerated species such as $^3$He are also picked up by the cyclotron damping mechanism and accelerated well beyond thermal energies. There are enough of them to possibly weaken the turbulence enough that it goes into the Sridhar-Goldreich regime, and at this point the situation would be sufficiently complex as to be beyond the scope of this paper.

In clusters of galaxies, the virial motion of galaxies can power Alfvenic turbulence. Considering the parameters $n= 10^{-2}\rm cm^{-3}$,  $B= 10^{-7}$ G,  $kT \sim 10$ keV. Then $v_A \sim 30$ km/s, the ion gyrofrequency is $ \omega_c \sim 10^{-3}\rm s^{-1}$, and the gyroradius is $r_g \sim 10^{11}$ cm.  Let us choose  $u^2 = 10^{-2} kT/m_p = [10^7$ cm s$^{-1}]^2$ $ \sim 10 v_A^2$.  The collisional mean free path is of  order $10^{22}$ cm, and the collision time of order $10^{14}$s. If turbulence can be excited on scale L much smaller than this, then the acceleration time is

\begin{equation}
\tau_{acc}= (l/r_g)^2 \omega_c^{-1} \sim 10^3 (L/10^{11}\rm cm)^2 s
\label{ratio}
\end{equation}
If the winds from individual stars of galaxies "rake" the intracluster medium and excite small scale turbulence at scale L,  then ions can be accelerated until their  gyroradii are comparable to L. For $L\sim 10^{16}$ cm,  this would imply that relativistic energies are attained -  enough for them to generate energetic electrons in nuclear collisions.

\section{Summary and Discussion}
We have gone back to an earlier work  (Eichler, 1979b) and reexamined it in light of the prediction of Sridhar and Goldreich (1995) that Alfven wave turbulence becomes tubular as it cascades to higher wavenumber. Tubular eddies aligned with the magnetic field (i.e. wavenumber nearly perpendicular to the field) are less efficient at accelerating particles via cyclotron resonance because resonant particles, having $v=\omega_c/k_{\parallel}$, cut across many tubes breadth-wise (i..e. many perpendicular wavelengths) during a single gyration. On the other hand, because the spectrum is a much steeper function of parallel wavenumber than in older theories (e.g. Kraichnan, 1965), the partially ionized heavy elements, having large gyroradii,  enjoy a stronger advantage (i.e. a shorter acceleration time) relative to lighter species with smaller gyroradii.  { We found the acceleration rate in Alfven turbulence with Sridhar-Goldreich geometry to be constant with energy in any one ion species, but with an acceleration rate that is proportional to the mass-to charge ratio.  This means that particles gain energy exponentially with time, and that, in the initial phases, the abundance enhancement of ultraheavies can be exponentially larger relative to lighter species.  It aslo suggests flat energy spectra until escape limits further acceleration so that the energy resides mostly in particles that have undergone many e-folds in energy.  It remains an open question whether the non-thermal tail of the spectrum reaches steady state, but, as the observed spectra are somewhat steeper  than we have calculated, maybe it doesn't. Rather, it is quite possible that the spectra we observe represent the superposition of particles from many acceleration sites, each with its own time-limited maximum energy, and this would lead to dramatic enhancement of partially ionized ultraheavies at the higher energies.}

We have shown that if strongly super-Alfvenic turbulence is stirred up by solar flares, then the Sridhar-Goldreich mode of cascade may set in at scales $l$ that are actually smaller than the thermal gyroradii of heavy ions, under the assumption that they move at the same "thermal" velocity at the lighter species (Eichler 1979a) due to other collisionless dissipative processes. In this case, the left hand side of equation (\ref{ratio}) can be large and some heavy ion species (e.g. oxygen, iron, which are sufficiently abundant that acceleration out to $\sim 1$ MeV/nucleon could in principle consume most of the energy budget) can damp the turbulence before it cascades to small enough scales to cyclotron resonate with protons, as discussed in E79. If such a situation were to obtain, it would further enhance the preferential acceleration of ultraheavies relative to lighter species.

 Observations suggest that impulsive SEP events do not coincide with coronal mass ejection, but rather originate at the interfaces between active regions and coronal holes (Wang et al., 2006). This is consistent with the picture that turbulent energy is exported from the strongly magnetized active region to the less magnetized coronal hole, where it is more super-Alfvenic, and thus less tubular.

It is still an  open question whether the heavier ion species are merely preferentially accelerated, as in a mostly  perpendicular cascade of turbulent energy, or whether they actually monopolize the energy budget of the cascading turbulence, which is more likely to be true for isotropic  than for tubular Alfven turbulence. Either way, the acceleration rate is a strongly increasing function of gyroradius, and ultraheavies can thus be  strongly enhanced relative to protons and other lighter species at high energies. However, the fact that rare ultraheavies ($Z\gg 26$) are enhanced even relative to iron  suggests that  monopolization of the energy budget, which is not likely in the case of rare ultraheavies,  is not the only mechanism for the enhancement. This woul disfavor simple isotropic Kolmogoroff or Kraichnan pictures of turbulence. 

A key observational quantity that might distinguish between these two distinct scenarios for the damping of a turbulent cascade is the fraction of flare energy that goes into energetic particles, as opposed to heat and/or mass ejection. Monopolization of the energy budget by superthermal heavy ions would predict that most of the cascading turbulence went into the superthermal particles, rather than heat.  However, other loss mechanisms, such as acoustic losses and non-linear Landau damping need to be considered before this matter can be fully settled. It is clear that any dissipation mechanism steepens the turbulence spectrum as it cascades to higher wavenumber, making the acceleration less efficient but increasing preference for high rigidity ion species.

It is therefore significant that observations of $^3$He-rich flares are consistent with a {\it low acceleration efficiency}. We may estimate that a coronal mass ejection (CME) requires at least $10^{11}$ g (Wang et al, 2006), hence at least $10^{26}$ ergs, to be detected, and CMEs that have been reported are typically more energetic than this by several orders of magnitude (Shimojo and  Shibata, 2000). By contrast, the fluence of SEPs, at energies of order 1 MeV,  is of order $10^3$ $^4$He nuclei per $cm^2$, which at 1 A.U. corresponds to an isotropic-equivalent energy of $\ll 10^{25}$ erg. The impulsive SEPs are apparently observed only when the field lines associated with the {\it boundaries} of active regions intersect the ecliptic - there is no indication that they are convected out of the corona, along with ejected mass, over a broad solid angle -  indicating that the particles escape only on isolated field lines, and that the true energy in escaping SEPs for these events is many orders of magnitude less than the isotropic-equivalent energy. 

The presence of some protons and $^4$He nuclei in the impulsive SEP population, is not unexpected. The enhancement is typically a sensitive but  smooth  function of mass to charge for all elements from He to ultraheavies, given reasonable assumptions about the charge states in a hot corona (Mason et al 2004). The enhancement  can be fitted as being proportional to $(M/Q)^{3.26}$. It may be that protons, which are not plotted in Mason (2004), are somewhat more abundant than simple extrapolation of  their formula (Reames, 1999). On the other hand, it could be that there is additional heating due to other mechanisms which competes with cyclotron damping when the rate of the latter is sufficiently small.

We have suggested here that SEP populations might shed light on how Alfven turbulence cascades and damps. At present, the picutre of
 relatively small energy content in SEPs, together with a systematic enhancement of ion species as a strong function of their charge to mass ratio,  is consistent with the Sridhar-Goldreich picture of tubular cascading of Alfven turbulence. More observational details are  needed to strengthen  this conclusion.

\acknowledgements
I thank  G. Mason for very helpful conversations and for encouraging a return to this work. I also thank Y. Lyubarsky and U. Keshet for helpful theoretical conversations. I gratefully acknowledge support from the Israel-U.S. Binational Science Foundation, the Israeli Science Foundation, and the Joan and Robert Arnow Chair of Theoretical Astrophysics. 

\bigskip
\bigskip

\centerline{References}

 \noindent Bogdan, T.J., Lee, M.A. Schneider, P. 1991, JGR, 96, 161

  \noindent Chandran, B.D.G., 2000, ApJ, 529, 513

  \noindent Eichler,  D. 1979a, ApJ, 229, 409

 \noindent Eichler, D., 1979b, ApJ, 229, 413

 \noindent  Eichler, D., 1979c, ApJ, 229, 419

  \noindent Fisk, L.A. 1978, ApJ 224, 1048 

 \noindent  Kennel, C.F. \& Engelman, F. 1966, Phys. Fluids, 9, 2377

  \noindent Kraichnan, R.H. 1965, Phys. of Fluids, 8, 1385

  \noindent Lee, M.A. \& Volk, H. J., 1975, ApJ, 198, 485
\bigskip

  \noindent Mason, G., Mazur, J.E., Dwyer, J.R., Jokipii, J.R., Gold, R.E., \&  Krimigis, S.M., (2004)   ApJ, 606, 555,

 \noindent  Miller, J.A. 1998, Space Science Reviews 86, 79,

\noindent Reames, D.V. 1999, Space Science Reviews, 90, 413

  \noindent Reames, D.V., 2000, ApJ, 540, L111

  \noindent Schlickeiser, R. \& Achatz, U. , 1993, J. Plasma Phys. 49,  63

  \noindent Shimojo, M. \& Shibata, K., 2000, ApJ 542, 1100 doi:10.1086/317024

 \noindent Sridhar, S. \& Goldreich, P., 1995, ApJ, 438, 753

 \noindent Wang, Y.M., Pick, M. \&  Mason, G.M. (2006) ApJ, 639, 495

   \end{document}